\documentclass[prd,superscriptaddress,showpacs,nofootinbib,amsmath,amssymb,aps,11pt]{revtex4}

\usepackage{bm}
\usepackage{amsfonts}
\usepackage{latexsym}
\usepackage[latin1]{inputenc}
\usepackage{graphicx}
\usepackage{amsmath}
\usepackage{palatino}
\usepackage{mathpazo}
\usepackage{booktabs}
\usepackage{dcolumn}

\def\jnl@style{\it}
\def\aaref@jnl#1{{\jnl@style#1}}

\def\aaref@jnl#1{{\jnl@style#1}}

\def\aj{\aaref@jnl{AJ}}                   % Astronomical Journal
\def\apj{\aaref@jnl{ApJ}}                 % Astrophysical Journal
\def\apjl{\aaref@jnl{ApJ}}                % Astrophysical Journal, Letters
\def\apjs{\aaref@jnl{ApJS}}               % Astrophysical Journal, Supplement
\def\apss{\aaref@jnl{Ap\&SS}}             % Astrophysics and Space Science
\def\aap{\aaref@jnl{A\&A}}                % Astronomy and Astrophysics
\def\aapr{\aaref@jnl{A\&A~Rev.}}          % Astronomy and Astrophysics Reviews
\def\aaps{\aaref@jnl{A\&AS}}              % Astronomy and Astrophysics, Supplement
\def\mnras{\aaref@jnl{Mon.~Not.~Roy.~Astron.~Soc.}}             % Monthly Notices of the RAS
\def\prd{\aaref@jnl{Phys.~Rev.~D}}        % Physical Review D
\def\prc{\aaref@jnl{Phys.~Rev.~C}}  % Physical Review C
\def\prl{\aaref@jnl{Phys.~Rev.~Lett.}}    % Physical Review Letters
\def\qjras{\aaref@jnl{QJRAS}}             % Quarterly Journal of the RAS
\def\skytel{\aaref@jnl{S\&T}}             % Sky and Telescope
\def\ssr{\aaref@jnl{Space~Sci.~Rev.}}     % Space Science Reviews
\def\zap{\aaref@jnl{ZAp}}                 % Zeitschrift fuer Astrophysik
\def\nat{\aaref@jnl{Nature}}              % Nature
\def\aplett{\aaref@jnl{Astrophys.~Lett.}} % Astrophysics Letters
\def\apspr{\aaref@jnl{Astrophys.~Space~Phys.~Res.}} % Astrophysics Space Physics Research
\def\physrep{\aaref@jnl{Phys.~Rep.}}      % Physics Reports
\def\physscr{\aaref@jnl{Phys.~Scr}}       % Physica Scripta
\def\commat{\aaref@jnl{Comm.~Math.~Phys.}}              % Communications in Mathematical Physics
\def\science{\aaref@jnl{Science}}               % Science
\def\cqg{\aaref@jnl{Classical Quant.~Grav.}}            % Classical and Quantum Gravity
\def\jpcs{\aaref@jnl{JPCS}}                                     % Journal of Physics Conference Series
\def\ijmpd{\aaref@jnl{Int.~J.~Mod.~Phys.~D}}                    % International Journal of Modern Physics D
\def\grg{\aaref@jnl{Gen.~Relat.~Gravit.}}               % General Relativity and Gravitation
\def\rpp{\aaref@jnl{Rep.~Prog.~Phys.}}          % Reports on Progress in Physics
\def\npa{\aaref@jnl{Nucl.~Phys.~A}}        % Nuclear Physics A
\def\lrr{\aaref@jnl{Living Rev.~Rel.}}                   % Living reviews in relativity
\def\jcap{\aaref@jnl{J.~Cosmology Astropart.~Phys.}}    % Journal of cosmology and astroparticle physics
\def\rmp{\aaref@jnl{Rev.~Mod.~Phys.}}   %Reviews of modern physics

\allowdisplaybreaks[1]
\begin{document}

\title{Neutron star solutions with curvature induced scalarization in the extended Gauss-Bonnet scalar-tensor theories }

\author{Daniela D. Doneva}
\email{daniela.doneva@uni-tuebingen.de}
\affiliation{Theoretical Astrophysics, Eberhard Karls University of T\"ubingen, T\"ubingen 72076, Germany}
\affiliation{INRNE - Bulgarian Academy of Sciences, 1784  Sofia, Bulgaria}

\author{Stoytcho S. Yazadjiev}
\email{yazad@phys.uni-sofia.bg}
\affiliation{Theoretical Astrophysics, Eberhard Karls University of T\"ubingen, T\"ubingen 72076, Germany}
\affiliation{Department of Theoretical Physics, Faculty of Physics, Sofia University, Sofia 1164, Bulgaria}

%%%%%%%%%%%%%%%%%%%%%%%%%%%%%%%%%%%%  DATE  %%%%%%%%%%%%%%%%%%%%%%%%%%%%%%%%%%%%
%\date{\today}

\begin{abstract}
In the present paper we study models of neutron stars in a class of extended scalar-tensor Gauss-Bonnet (ESTGB) theories
for which the scalar degree of freedom is exited only in the strong curvature regime. We show that in the framework
of the ESTGB theories under consideration there exist new neutron star solutions which are formed via spontaneous scalarization 
of the general relativistic neutron stars. In contrast to the spontaneous scalarization in the standard scalar-tensor theories
which is induced by the presence of matter, in our case the scalarization is induced by the spacetime curvature.           
\end{abstract}


\maketitle

\section{Introduction}

Among the most natural generalizations of Einstein's general relativity are the extended scalar-tensor theories (ESTT) of gravity. 
In these theories the usual Einstein-Hilbert action is suplimented with all possible algebraic 
curvature invariants of second order and with a dynamical scalar field nonminimally coupled to these invariants \cite{Berti_2015}-\cite{Pani_2011a}. A particular  sector of the ESTT is the extended scalar-tensor-Gauss-Bonnet (ESTGB) gravity for which 
the scalar field is coupled exactly to the Gauss-Bonnet invariant.  An important property of the ESTGB gravity
is that the field equations are of second order as in general relativity (GR) and the theory is free from ghosts.
In the present paper  we shall focus namely  on the ESTGB gravity.

Neutron stars are the perfect laboratories for testing the modified theories of gravity. That is why it is natural to study 
the neutron stars in the  ESTGB gravity. Neutron stars in particular  class of  the ESTGB gravity (characterized  
by the coupling function $f(\varphi)=\alpha e^{\beta\varphi}$ with $\alpha$ and $\beta$ being constants), the so-called Einstein-dilaton-Gauss-Bonnet (EdGB) gravity, have attracted  interest in the recent years. Neutron stars in the EdGB gravity were constructed for the first time in \cite{Pani_2011b} including both the static and the slowly rotating cases. It was shown in \cite{Pani_2011b} that the maximum mass decreases with respect to pure general relativity and for fixed values of the coupling parameter no neutron star solutions exist above some critical central energy density. This can serve as a way to impose constraint on the theory in the future when the equation of state uncertainties are reduced one should simply require that the coupling parameter is below some critical values chosen in such a way that the maximum mass is above the two solar mass barrier. The studies show that the neutron star moment of inertia in EdGB gravity deviate from the one in pure Einstein's theory by a few percents at most, that can hardly be observed. The axial quasinormal modes of neutron stars in EdGB were examined in \cite{Slasedo_2016} and an interesting observation is that the universal (equation of state independent) relations for the oscillation modes that are available in Einstein's theory still hold for the EdGB gravity. Rapidly rotating neutron stars in EdGB gravity were constructed in \cite{Kleinhaus_2014} and \cite{Kleinhaus_2016} where also the universal relations between the normalized moment of inertia and quadrupole moment were examined and it was shown that the deviations from pure general relativity are small in these relations.

In the present paper we shall consider a class of ESTGB theories with a scalar coupling functions for which the scalar degree of freedom is excited only in the strong curvature regime like the class studied in \cite{Doneva_2017} (see also  \cite{Silva_2017} and \cite{Antoiou_2017a}-\cite{Antoiou_2017b} ). In particular we shall show that in the mentioned class of ESTGB theories there exists new neutron star solutions which are formed by spontaneous scalarization of the general relativistic neutron stars in the extreme curvature regime. In contrast with the standard spontaneous scalarization \cite{Damour_1993} which is induced by the presence of matter, in our case the scalarization is induced by the curvature of the spacetime.  During the preparation of this manuscript, a preprint studying a similar 
neutron star models  appeared on arXiv \cite{Silva_2017}. It has to be noted however that neutron stars in  \cite{Silva_2017} were studied 
only perturbatively, i.e. on the level of eq.(\ref{PESF}). Contrary to \cite{Silva_2017} we study the neutron stars in the mention class of ESTGB gravity by exploring the full nonlinearity of the field equations.

\section{Basic equations and setting the problem}
We consider the  ESTGB theories in the presence of  matter. In this case the action of the ESTGB theories is the following:  

\begin{eqnarray}
S=&&\frac{1}{16\pi}\int d^4x \sqrt{-g} 
\Big[R - 2\nabla_\mu \varphi \nabla^\mu \varphi  
+ \lambda^2 f(\varphi){\cal R}^2_{GB} \Big] + S_{\rm matter}  (g_{\mu\nu},\chi) ,\label{eq:quadratic}
\end{eqnarray}
where $R$ denotes the Ricci scalar with respect to the spacetime metric $g_{\mu\nu}$, $\nabla_{\mu}$ is the covariant derivative with respect to the spacetime metric $g_{\mu\nu}$ and  $f(\varphi)$ is the coupling function for the scalar field $\varphi$.  The Gauss-Bonnet coupling constant $\lambda$ has  dimension of $length$ and ${\cal R}^2_{GB}$ denotes the Gauss-Bonnet invariant defined by ${\cal R}^2_{GB}=R^2 - 4 R_{\mu\nu} R^{\mu\nu} + R_{\mu\nu\alpha\beta}R^{\mu\nu\alpha\beta}$ where $R_{\mu\nu}$ and $R_{\mu\nu\alpha\beta}$ are  the Ricci tensor and  the Riemann tensor respectively. $S_{\rm matter}$ is the matter action where the matter fields are collectively denoted by $\chi$. 

The field equations derived from the action are 
\begin{eqnarray}\label{FE}
&&R_{\mu\nu}- \frac{1}{2}R g_{\mu\nu} + \Gamma_{\mu\nu}= 2\nabla_\mu\varphi\nabla_\nu\varphi -  g_{\mu\nu} \nabla_\alpha\varphi \nabla^\alpha\varphi  + 8\pi T^{\rm matter}_{\mu\nu},\\
&&\nabla_\alpha\nabla^\alpha\varphi= -  \frac{\lambda^2}{4} \frac{df(\varphi)}{d\varphi} {\cal R}^2_{GB},
\end{eqnarray}
where  $T^{\rm matter}_{\mu\nu}$ is the matter energy momentum tensor. $\Gamma_{\mu\nu}$ is defined by 

\begin{eqnarray}
\Gamma_{\mu\nu}&=& - R(\nabla_\mu\Psi_{\nu} + \nabla_\nu\Psi_{\mu} ) - 4\nabla^\alpha\Psi_{\alpha}\left(R_{\mu\nu} - \frac{1}{2}R g_{\mu\nu}\right) + 
4R_{\mu\alpha}\nabla^\alpha\Psi_{\nu} + 4R_{\nu\alpha}\nabla^\alpha\Psi_{\mu} \nonumber \\ 
&& - 4 g_{\mu\nu} R^{\alpha\beta}\nabla_\alpha\Psi_{\beta} 
+ \,  4 R^{\beta}_{\;\mu\alpha\nu}\nabla^\alpha\Psi_{\beta} 
\end{eqnarray}  
with 

\begin{eqnarray}
\Psi_{\mu}= \lambda^2 \frac{df(\varphi)}{d\varphi}\nabla_\mu\varphi .
\end{eqnarray}

Using  the contracted Bianchi identies and the field equations, after some algebra one can show that the matter energy-momentum tensor  
$T^{\rm matter}_{\mu\nu}$ satisfies 

\begin{eqnarray}\label{BID}
\nabla^{\mu}T^{\rm matter}_{\mu\nu}=0. 
\end{eqnarray}

Since the purpose of the present paper is to study the structure of static and spherically symmetric neutron stars in
the ESTGB theories, we consider a static and spherically  symmetric spacetime with a metric  

\begin{eqnarray}
ds^2= - e^{2\Phi(r)}dt^2 + e^{2\Lambda(r)} dr^2 + r^2 (d\theta^2 + \sin^2\theta d\phi^2 )
\end{eqnarray}   
and the matter source to be a perfect fluid with $T^{\rm matter}_{\mu\nu}=(\rho + p)u_{\mu}u_{\nu} + pg_{\mu\nu}$ 
where $\rho$, $p$ and $u^{\mu}$ are the energy density, presure and 4-velocity of the fluid, respectively. We also require the perfect  
fluid and the scalar field to respect the staticity and the spherical symmetry. With these
conditions imposed, the dimensionally reduced field equations are:

\begin{eqnarray}
&&\frac{2}{r}\left[1 +  \frac{2}{r} (1-3e^{-2\Lambda})  \Psi_{r}  \right]  \frac{d\Lambda}{dr} + \frac{(e^{2\Lambda}-1)}{r^2} 
- \frac{4}{r^2}(1-e^{-2\Lambda}) \frac{d\Psi_{r}}{dr} - \left( \frac{d\varphi}{dr}\right)^2=8\pi \rho e^{2\Lambda}, \label{DRFE1}\\ && \nonumber \\
&&\frac{2}{r}\left[1 +  \frac{2}{r} (1-3e^{-2\Lambda})  \Psi_{r}  \right]  \frac{d\Phi}{dr} - \frac{(e^{2\Lambda}-1)}{r^2} - \left( \frac{d\varphi}{dr}\right)^2=8\pi p e^{2\Lambda},\label{DRFE2}\\ && \nonumber \\
&& \frac{d^2\Phi}{dr^2} + \left(\frac{d\Phi}{dr} + \frac{1}{r}\right)\left(\frac{d\Phi}{dr} - \frac{d\Lambda}{dr}\right)  + \frac{4e^{-2\Lambda}}{r}\left[3\frac{d\Phi}{dr}\frac{d\Lambda}{dr} - \frac{d^2\Phi}{dr^2} - \left(\frac{d\Phi}{dr}\right)^2 \right]\Psi_{r} 
\nonumber \\ 
&& \hspace{0.5cm} - \frac{4e^{-2\Lambda}}{r}\frac{d\Phi}{dr} \frac{d\Psi_r}{dr} + \left(\frac{d\varphi}{dr}\right)^2=8\pi p e^{2\Lambda}, \label{DRFE3}\\ && \nonumber \\
&& \frac{d^2\varphi}{dr^2}  + \left(\frac{d\Phi}{dr} \nonumber - \frac{d\Lambda}{dr} + \frac{2}{r}\right)\frac{d\varphi}{dr} \nonumber \\ 
&& \hspace{0.5cm} - \frac{2\lambda^2}{r^2} \frac{df(\varphi)}{d\phi}\Big\{(1-e^{-2\Lambda})\left[\frac{d^2\Phi}{dr^2} + \frac{d\Phi}{dr} \left(\frac{d\Phi}{dr} - \frac{d\Lambda}{dr}\right)\right]    + 2e^{-2\Lambda}\frac{d\Phi}{dr} \frac{d\Lambda}{dr}\Big\} =0, \label{DRFE4}
\end{eqnarray}
where  

\begin{eqnarray}
\Psi_{r}=\lambda^2 \frac{df(\varphi)}{d\varphi} \frac{d\varphi}{dr}.
\end{eqnarray}

In addition to the above dimensionally reduced field equations we have to add the equation for the hydrostatic equilibrium of the fluid

\begin{eqnarray}\label{FEE}
\frac{dp}{dr} = - (\rho + p) \frac{d\Phi}{dr},
\end{eqnarray}  
which follows from eq.(\ref{BID}).

In the present paper we consider a class of ESTGB theories with coupling functions $f(\varphi)$ that fulfill the condition $\frac{df}{d\varphi}(0)=0$. Without loss of generality we also can impose   $\frac{d^2f}{d\varphi^2}(0)=\epsilon$ where $\epsilon=\pm 1$. 

Using the expansion of the field equations close to the center of the star one can derive important relations determining the existence of solutions. If one substitutes
\begin{equation}
\Lambda=\Lambda_0 + \Lambda_1 r+\frac{1}{2} \Lambda_2 r^2  + O(r^3)\;\;\;\Phi=\Phi_0+\Phi_1 r+\frac{1}{2} \Phi_2 r^2 + O(r^3) \;\;\; \varphi=\varphi_0+\varphi_1 r + \frac{1}{2} \varphi_2 r^2 + O(r^3)
\end{equation}
and expands the differential equations \eqref{DRFE1}--\eqref{DRFE4} around the center of the star then it is easy to obtain that $\Lambda_0=0$, $\Lambda_1=0$, $\Phi_1=0$ and $\varphi_1=0$. In addition, the following condition should be also fulfilled
\begin{equation}\label{eq:SolutionConstrain}
9  \lambda^4  \left(\frac{df}{d\varphi}(\varphi_0)\right)^2 (\Lambda_2)^4 - 72 \pi \lambda^4 p_0  \left(\frac{df}{d\varphi}(\varphi_0)\right)^2   (\Lambda_2)^3 - 6\pi \rho_0  \Lambda_2 + 16 \pi^2 \rho_{0}^2=0.
\end{equation}
This is a fourth order algebraic equation for $\Lambda_2$ that depends on the values of the pressure and energy density at the center of the star $p_0$ and $\rho_0$, as well as the central value of the scalar field $\varphi_0$.

Due to the particular form of the coupling function, i.e. $\frac{df}{d\varphi}(0)=0$, one can easily check that the case with $\varphi=0$ is always a solution of the field equations and it coincides with the pure general relativistic solution. The question is whether additional solutions with nontrivial scalar field exist for a certain range of parameters. One can straightforward make a parallel with the spontaneous scalarization of neutron stars in scalar-tensor theories \cite{Damour_1993} where the scalar field is sourced by the matter. Our case is different, though, since the energy-momentum tensor does not enter in the field equation for the scalar field \eqref{FE}. Instead the scalar field is sourced by the spacetime curvature via the Gauss-Bonnet invariant. In order to gain intuition under what circumstances such scalarized solutions exist one can first check the stability of the trivial solution with $\varphi=0$. For this purpose we will derive the perturbation of the pure general relativistic solution with $\varphi=0$ within the framework of the described class of ESTGBT. It is easy to check that the equations for the perturbations of the metric decouple from the equation for the perturbation of the scalar field, and they are the same as in pure general relativity. That is why it is enough to consider only the perturbation of the scalar field that is governed by the following equation
\begin{eqnarray}\label{PESF}
\Box_{(*)} \delta\varphi + \frac{1}{4}\lambda^2  {\cal R}^2_{GB(*)} \delta\varphi=0, 
\end{eqnarray} 
where $\Box_{(*)}$ and ${\cal R}^2_{GB(*)}$ are the D'alambert operator and the Gauss-Bonnet invariant for 
the pure general relativistic solution. Since we are dealing with static and spherically symmetric solutions the variables can be separated
\begin{eqnarray}
\delta\varphi= \frac{u(r)}{r} e^{-i\omega t}Y_{lm}(\theta,\phi),
\end{eqnarray}
where $Y_{lm}(\theta,\phi)$ are the standard spherical harmonics.  Thus we obtain the following perturbation equation
\begin{eqnarray}\label{eq:PertEq}
e^{\Phi_*-\Lambda_*} \frac{d}{dr}\left[e^{\Phi_*-\Lambda_*} \frac{d}{dr}u(r)\right] 
+ \left[\omega^2- U(r)\right]u(r)=0,
\end{eqnarray}
where $\Phi_*$ and $\Lambda_*$ are the metric function of the compact star solution in the pure general relativistic case. Substituting $\frac{d}{dr_*}=e^{\Phi_*-\Lambda_*} \frac{d}{dr}$ would obviously cast the above equation in the well known Schroedinger's form and the corresponding potential $U(r)$ is given by 
\begin{equation} \label{eq:Potential}
U(r)=e^{2\Phi_*}\left[\frac{e^{-2\Lambda_*}}{r}(\frac{d\Phi_*}{dr}-\frac{d\Lambda_*}{dr}) +\frac{l(l+1)}{r^2}-\lambda^2\epsilon {\cal R}^2_{GB(*)}\right].
\end{equation}
The Gauss-Bonnet invariant  for the static spherically symmetric purely generally relativistic case in the presence of matter is 
\begin{equation}
{\cal R}^2_{GB(*)}=\frac{12m_{*}^2}{r^6}-\frac{32\pi}{r^3}(2\pi p_* r^3 + m_*)\rho_*
\end{equation}
where the local mass is defined as
\begin{equation}
m_*=\frac{r}{2}\left(1-e^{-2\Lambda_*}\right)
\end{equation}
and we recall that $\epsilon=\frac{d^2 f}{d\varphi^2}(0)$.

If the potential $U(r)$ has a substantial negative part then the eigenvalues $\omega^2$ could become negative indicating the appearance of instability of the pure general relativistic solution. It is natural to expect that in this case a new branch of solutions with nontrivial scalar field would bifurcate from the pure general relativistic one and we will construct such solutions explicitly in the next section.

\section{Numerical setup and results}

In the present paper we will first work  with the following coupling function  
\begin{equation}
f(\varphi)=  \frac{1}{2\beta} \left[\exp(-\beta\varphi^2)-1\right] \label{eq:coupling_function},
\end{equation} 
where $\beta>0$ is a parameter and $\epsilon=-1$.

In the numerical results, presented below, we use the dimensionless parameter
\begin{eqnarray}
\lambda \to \frac{\lambda}{R_0}
\end{eqnarray}
where $R_0 = 1.47664$ km which corresponds to one solar mass.

The system of reduced field equations \eqref{DRFE1}--\eqref{FEE} is solved numerically with the following boundary conditions -- asymptotic flatness at infinity and regularity at the center of the star:
\begin{eqnarray}
\Lambda|_{r\rightarrow\infty} \rightarrow 0,\;\; \Phi|_{r\rightarrow\infty} \rightarrow 0, \;\; \varphi|_{r\rightarrow\infty} \rightarrow 0\;\;.   \label{eq:NS_inf}
\end{eqnarray} 
and
\begin{eqnarray}
\Lambda|_{r\rightarrow 0} \rightarrow 0, \;\; \left.\frac{d\Phi}{dr}\right|_{r\rightarrow 0} \rightarrow 0, \;\; \left.\frac{d\varphi}{dr}\right|_{r\rightarrow 0} \rightarrow 0\;\;.   \label{eq:NS_0}
\end{eqnarray} 
This defines a boundary value problem that is solved with a shooting method where the shooting parameters are the values of the scalar field $\varphi_0$ and the metric function $\Phi_0$ at the center of the star. In our numerical simulations we have chosen to work with the local mass $m(r)$ defined as $m=\frac{r}{2}\left(1-e^{-2\Lambda}\right)$ instead of $\Lambda$. 

The mass of the neutron star $M$ is just the asymptotic value of $m(r)$ at infinity while the dilaton charge $D$ is obtained through the asymptotic of the scalar field $\varphi$: 
\begin{equation}
\varphi\approx \frac{D}{r} + O(1/r^2).
\end{equation}

As we commented, for the class of ESTGB theories under consideration, the case with $\varphi=0$ is always a solution of the field equations. This coincides with the pure general relativistic solutions and we will call then the trivial solutions. In our analysis we will be searching for bifurcation points of the trivial branch of solutions where new branches of solutions with nontrivial scalar field (nontrivial solutions) appear. It is numerically difficult to gain intuition about these bifurcation points by solving the field equations \eqref{DRFE1}--\eqref{FEE} since we have nonuniqueness of the solutions and the shooting procedure does not easily converge to a nontrivial solution. Moreover, the conditions for the existence of a real root of equation \eqref{eq:SolutionConstrain} can be easily violated for a large range of parameters. Thus, similar to the black hole case \cite{Doneva_2017}, we have determined beforehand the points of bifurcation using the perturbation equation \eqref{eq:PertEq} with the following boundary conditions -- at the center of the star $du/dr|_{(r\rightarrow 0)}=0$ and $u(r)$ should be finite at $r\rightarrow 0$, while the eigenvalue $\omega^2$ is determined by the requirement that $u(r)|_{r\rightarrow \infty}=0$. More precisely, the bifurcations appear at the points where the eigenvalues of \eqref{eq:SolutionConstrain} become zero. Naturally, more than one bifurcation point can exist depending on how big the negative part of the potential defined by eq. \eqref{eq:Potential}. 

We have employed a realistic equation of state MPA1 \cite{Muther1987} but in the numerical solutions we have used its piecewise polytropic approximation form \cite{Read2009}. This equation of state produces maximum masses well above two solar masses and it is also in agreement with the constraints set by the observation of double neutron star mergers \cite{LIGO_NSMerger}.
 
The presented results are verified numerically in several different ways. First we have compared our code against the results presented in \cite{Pani_2011b} when employing the specific coupling function used there. Second, the positions of the bifurcation points calculated using the perturbation equation \eqref{eq:PertEq} and obtained from the direct solution of the field equations \eqref{DRFE1}-\eqref{FEE}, coincide. Moreover, the mass of the neutron star can be obtained either by the asymptotic of the local mass $m(r)$ or the metric function $\Phi(r)$ and the two results are in a very good agreement. At the end, we have proven that indeed the solutions disappear (i.e. the branches are terminated) at points where the condition for the existence of real roots of the fourth order algebraic equation for $\Lambda_2$ -- eq. \eqref{eq:SolutionConstrain}, is violated.

\begin{figure}[htb]
	\includegraphics[width=0.3\textwidth]{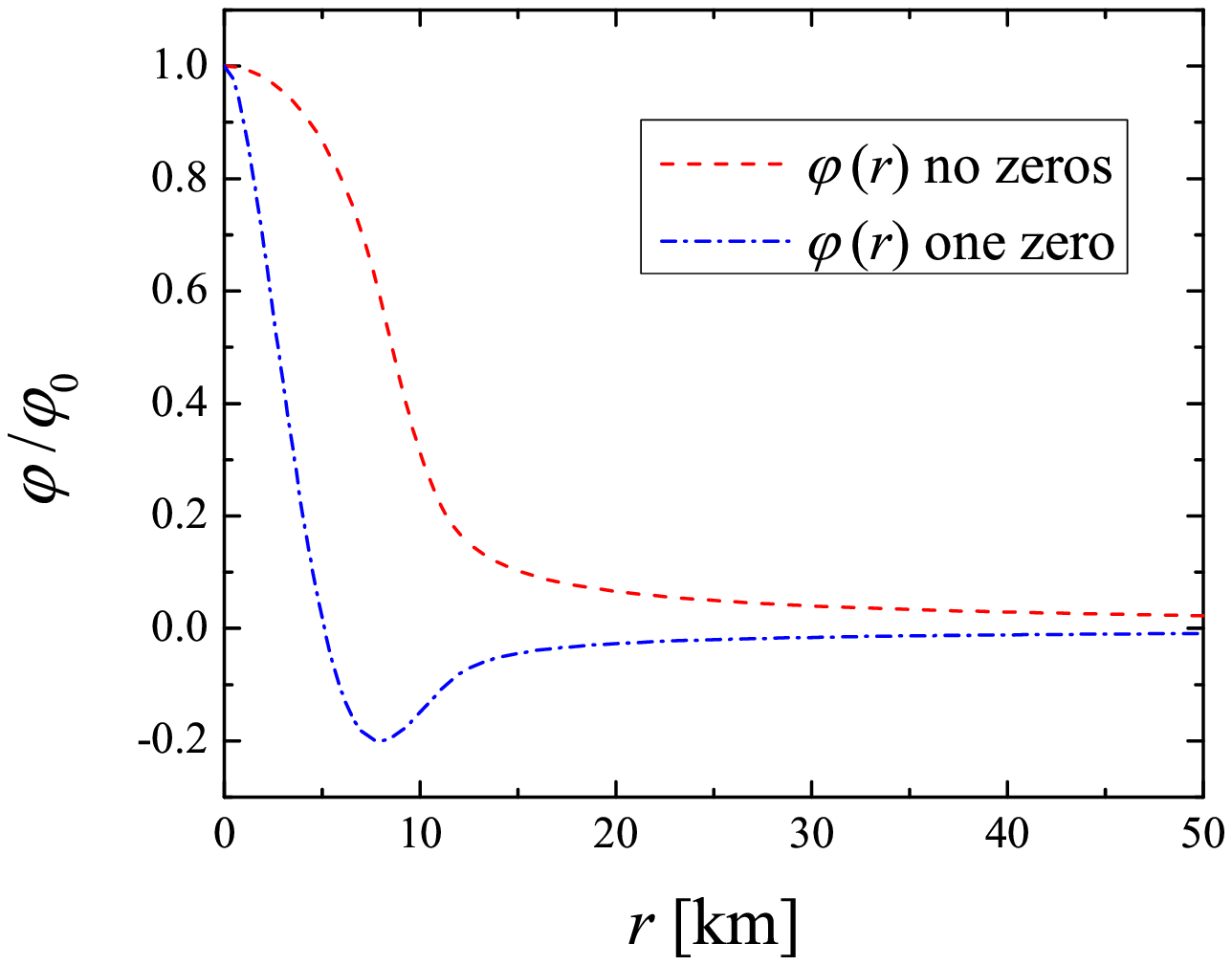}
	\includegraphics[width=0.3\textwidth]{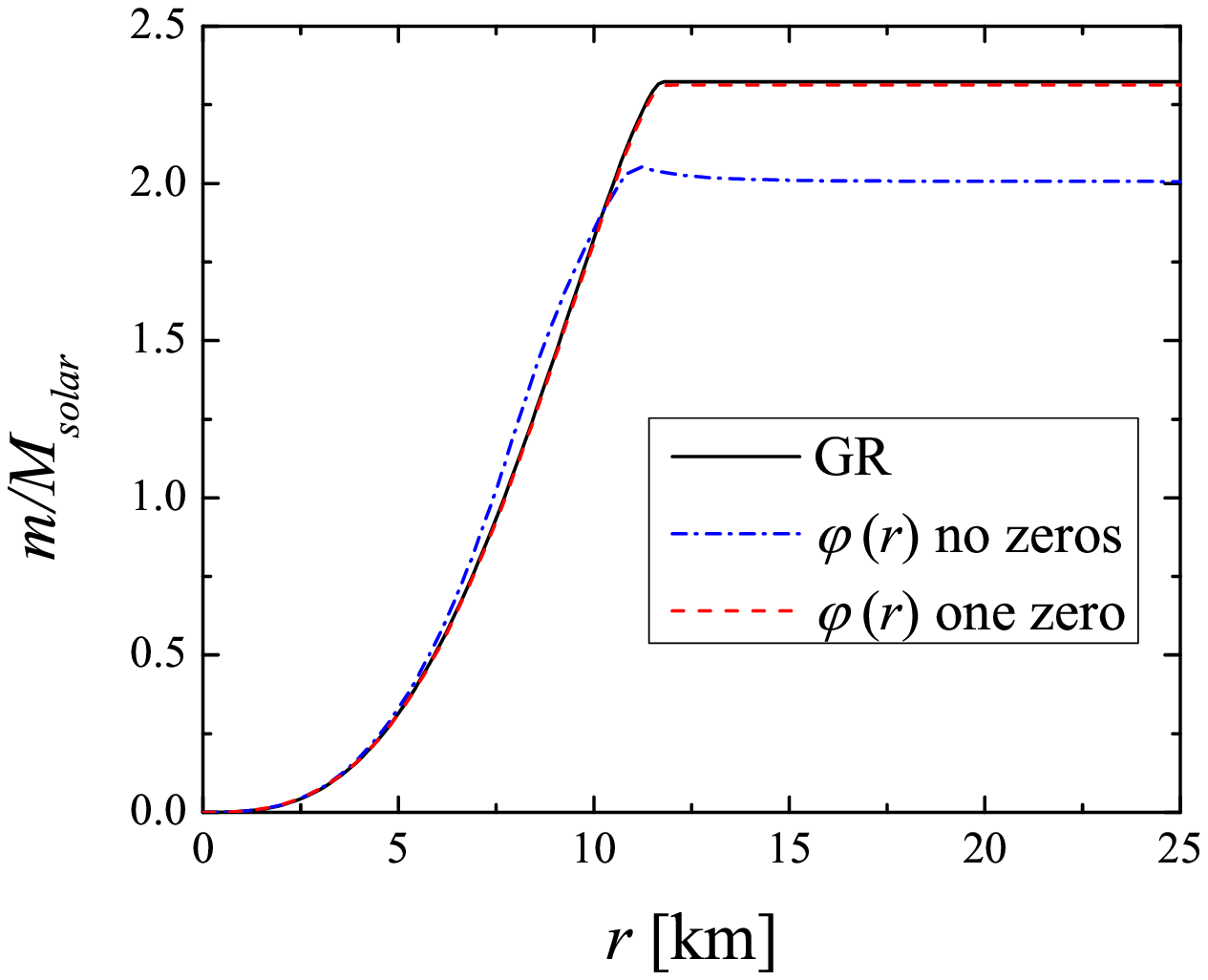}
	\includegraphics[width=0.3\textwidth]{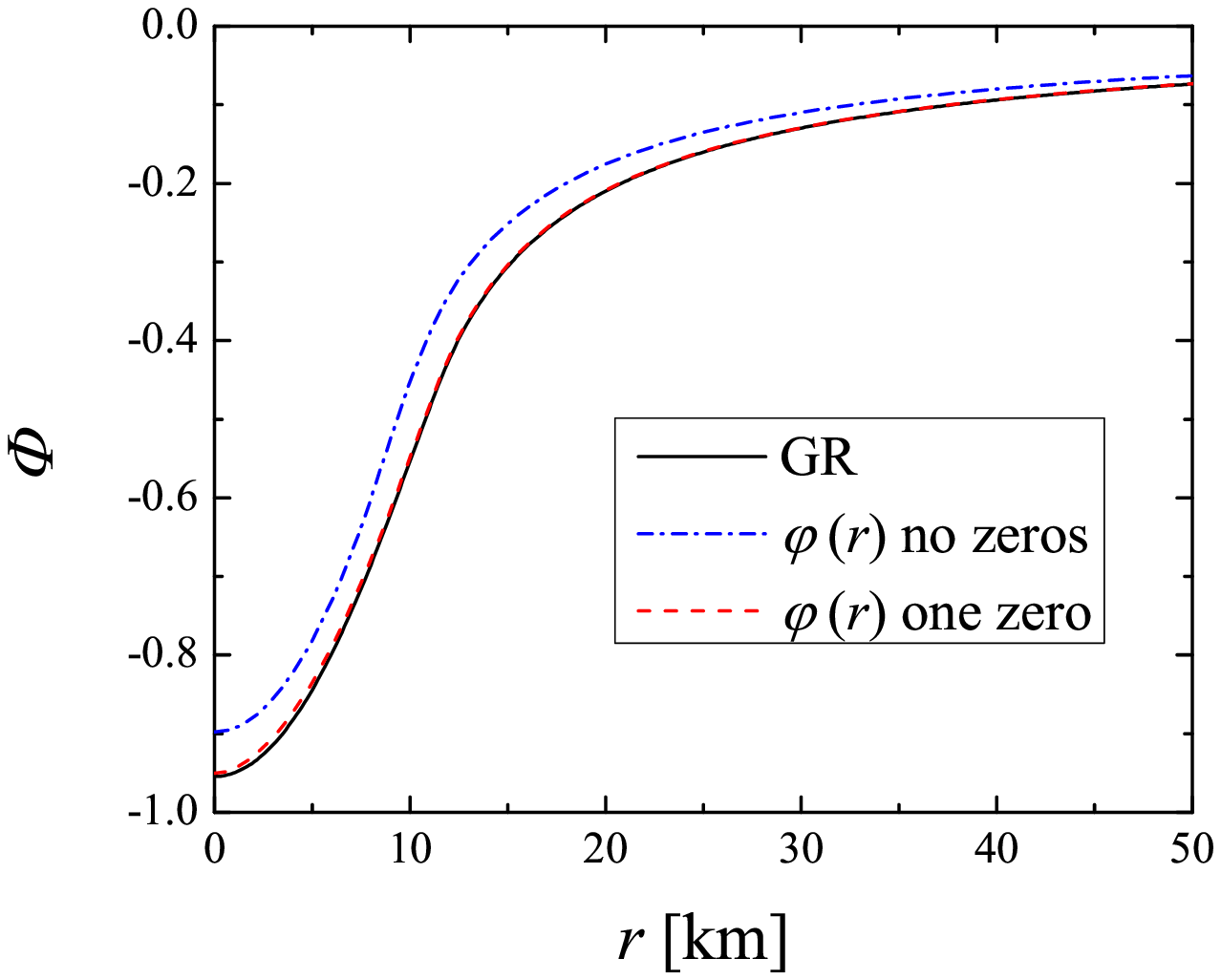}
	\caption{The normalized scalar field $\varphi/\varphi_0$ (left panel) , where $\varphi_0$ is the value of the scalar field at the center of the star for the corresponding solution, and the metric functions $m$ (middle panel) and $\Phi$ (right panel) as functions of the distance from the center of the star. We have fixed $\lambda=20$ and $\beta=200$ and all solutions have the same central energy density $\rho_0=1.33 \times 10^{15} {\rm g/cm^3}$. As one can see three different solutions exist -- one with trivial scalar field $\varphi=0$, one with scalar field that has no zeros and one with scalar field that has one zero, which demonstrates the observed nonuniqueness of the solutions.}
	\label{fig:phi0_m_Phi_r}
\end{figure}

The dependence of the metric functions and the scalar field\footnote{Since the model under consideration is invariant under the symmetry $\varphi\to -\varphi$ we shall present only the results for $\varphi_{0}>0$.} on the radial coordinate $r$ is shown in Fig. \ref{fig:phi0_m_Phi_r} for some representative solutions. We have chosen $\lambda=20$ and $\beta=200$ and all the solutions have central energy density $\rho_0=1.33 \times 10^{15} {\rm g/cm^3}$. The value of $\rho_0$ is chosen in such a way that three solutions exist -- the trivial one with zero scalar field (the pure general relativistic solution), a solution with no zeros of the scalar field and a solution with one zero of the scalar field. The scalar field plotted in the left panel is normalized to the corresponding value of the scalar field at center of the star -- $\varphi_0 = 0.116$ for the solution with no zeros and $\varphi_0 = 0.017$ for the solution with one zero. Given the fact that $\varphi_0$ is almost one order of magnitude smaller in the latter case, it is natural to expect that the metric functions of the solution with one zero would be much closer to the pure general relativistic case compared to the solution with no zero. As a matter of fact the maximum $\varphi_0$ that can be achieved before the condition \eqref{eq:SolutionConstrain} is violated for the branch of solutions with one zero is very close to the value given above. Therefore, this branch is very short and deviates from the pure general relativistic case only marginally and that is why we have not plotted it in the rest of the figures. Moreover, it is expected that only the branch where $\varphi(r)$ has no zeros is stable while the rest are unstable as commented below.

\begin{figure}[htb]
	\includegraphics[width=0.495\textwidth]{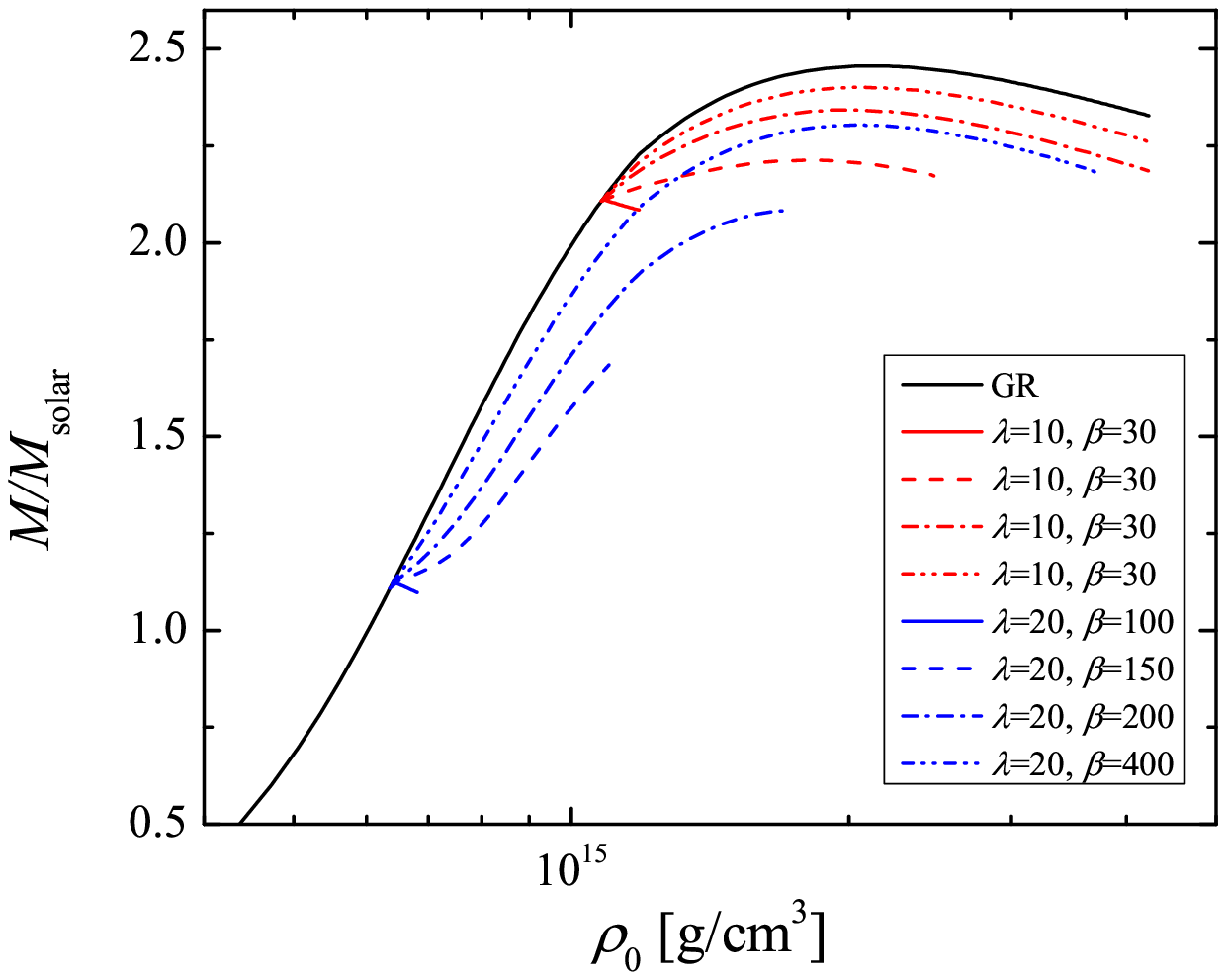}
	\includegraphics[width=0.495\textwidth]{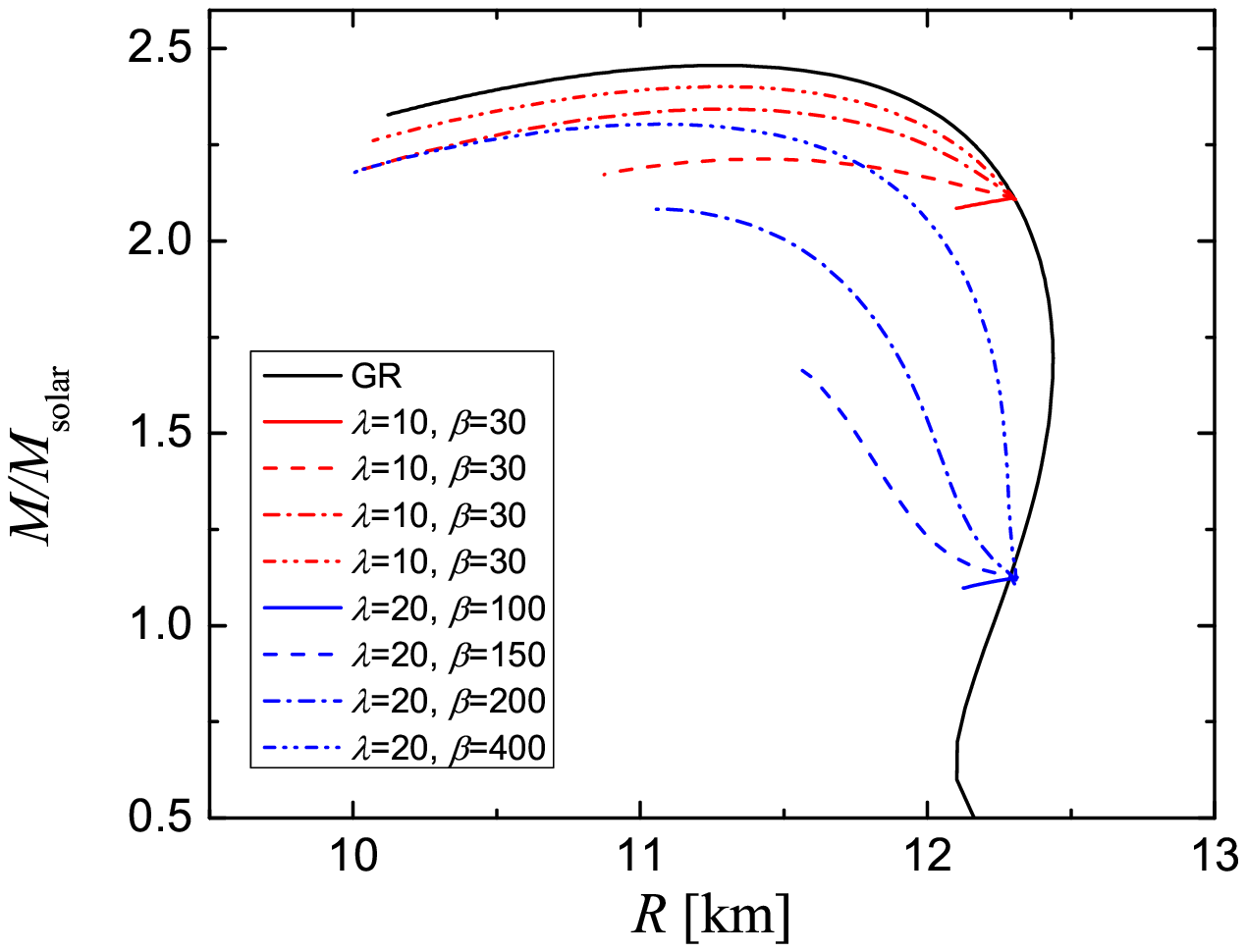}
	\caption{The mass as a function of the central energy density (left panel) and as a function of the radius (right panel). The case of pure general relativity (trivial branch with $\varphi=0$) is shown as well as several scalarized branches with different values of $\lambda$ and $\beta$. Only solutions characterized by no zero of the scalar field function $\varphi(r)$ are shown. }
	\label{fig:M_R_eps}
\end{figure}

The mass as a function of the central energy density  and of the radius is shown in Fig. \ref{fig:M_R_eps} for both the trivial solution (the pure GR case) and the scalarized solutions for different values of the parameters $\lambda$ and $\beta$. The branches terminate either at a fixed central energy density chosen to be $\rho_0=4.2 \times 10^{15} {\rm g/cm^3}$ or at the point where the condition for the existence of real roots of equation \eqref{eq:SolutionConstrain} is violated. According to eq. \eqref{eq:PertEq} the position of the bifurcation point is controlled only by the value of $\lambda$ for the particular choice of the coupling function \eqref{eq:coupling_function}. Only solutions characterized by no zero of the scalar field function $\varphi(r)$ are shown in the figure and that is why two bifurcation points exit for two different values of $\lambda$. Of course for every $\lambda$ other bifurcation points can exit that are characterized by scalarized solution with one or more zeros of the scalar field and our observation is that this happens at larger central energy densities $\rho_0$. As commented above, though, these branches are in general much shorter and they deviate from the general relativistic case almost negligibly. That is why we have not presented them in this figure. 

As one can see the maximum mass decreases for the scalarized solutions and in the limit when $\beta \rightarrow \infty$ the solutions tend to the trivial ones with $\varphi=0$. This poses limits on the free parameters, namely for a given equation of state these parameters should be such that the maximum mass is above two solar masses. Setting limits on $\lambda$ alone in this way is not possible since for every $\lambda$ we can choose a large enough $\beta$ so that the maximum mass is close to the pure general relativistic case (assuming the equation of state we work with gives above two solar masses in the $\varphi=0$ case). On the other hand, once $\lambda$ is fixed there exist a minimum $\beta$ below which two solar masses can not be achieved. Deriving a relation for the threshold $\beta$ is not so straightforward since this would depend heavily on the equation of state. Moreover the maximum mass is not always the turning point of the branch, instead the branch might be terminated before reaching the turning point because the condition for the existence of real roots of equation \eqref{eq:SolutionConstrain} is violated.

\begin{figure}[htb]
	\includegraphics[width=0.495\textwidth]{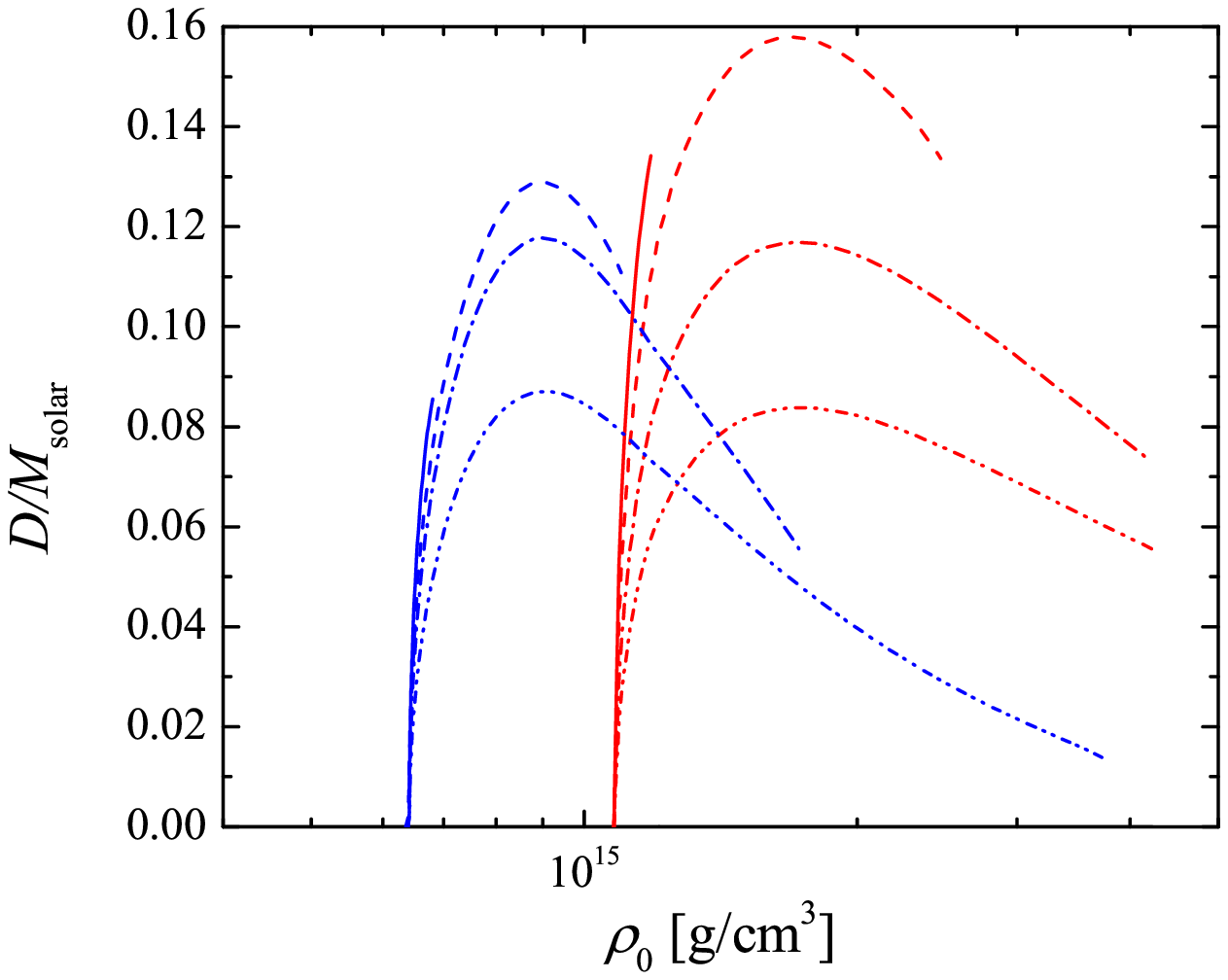}
	\includegraphics[width=0.495\textwidth]{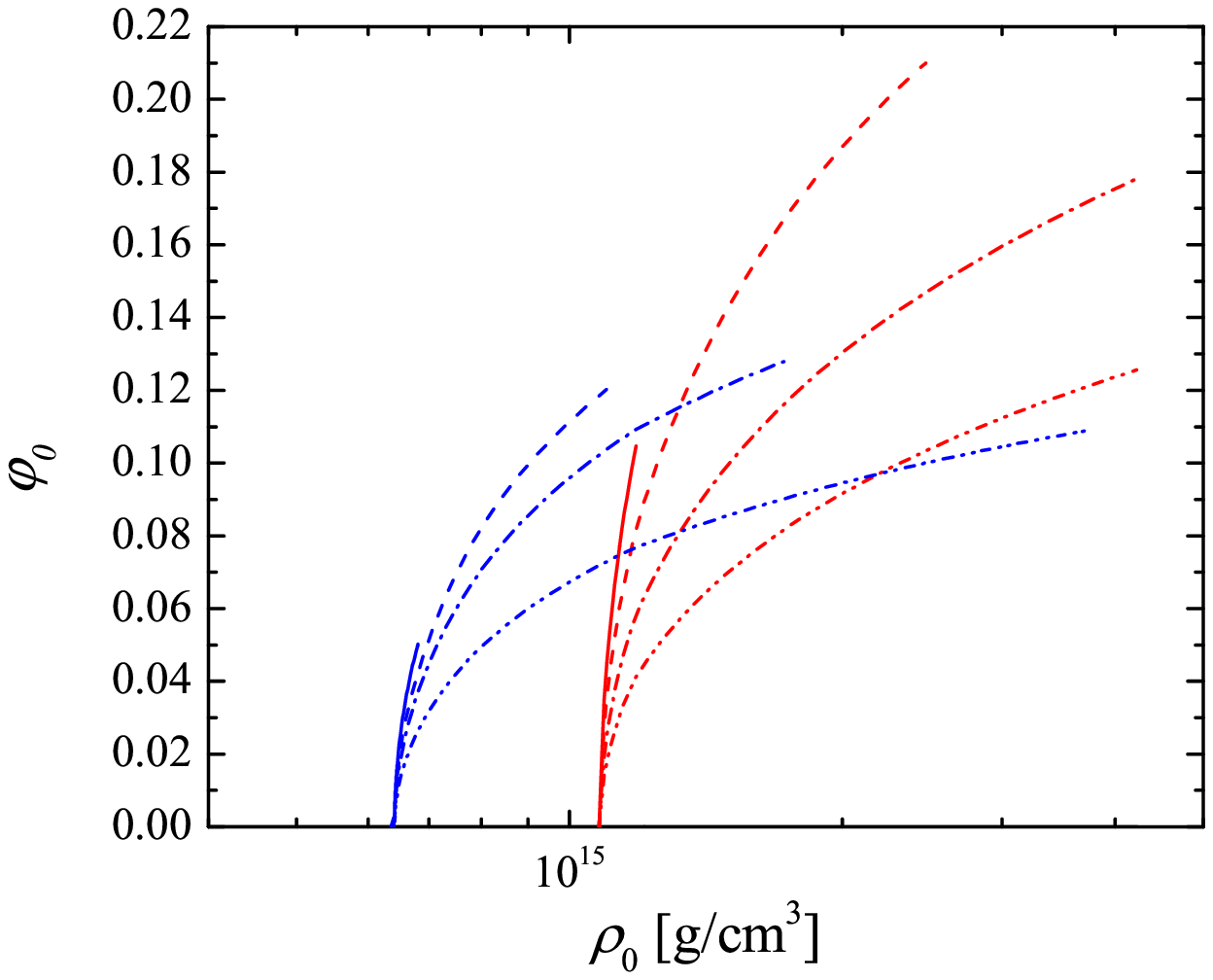}
	\caption{The dilaton change (left panel)  and the scalar field at the center of the star (right panel) as functions of the central energy density. The notations are the same as in Fig. \ref{fig:M_R_eps}}
	\label{fig:D_phi0_eps}
\end{figure}

The scalar charge and the central value of the scalar field are shown in Fig. \ref{fig:D_phi0_eps} as functions of the central energy density. As one can see, while the dilaton change first increase after the bifurcation point and afterwards decreases, the central value of the scalar field is always a monotonic function of the central energy density. 

\begin{figure}[htb]
	\includegraphics[width=0.495\textwidth]{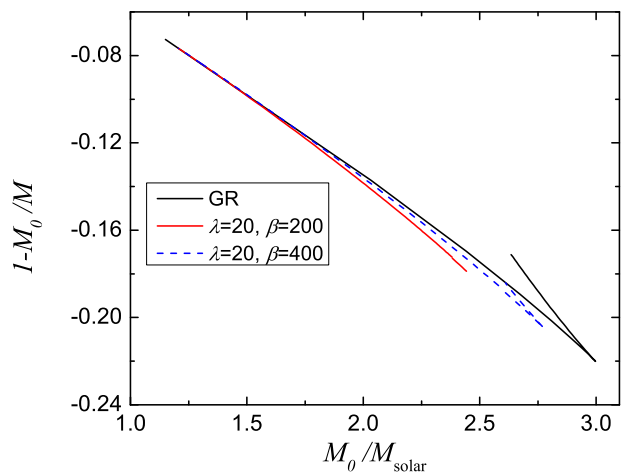}
	\caption{The quantity $1-M_0/M$, that is related to the binding energy of the star, as a function of $M_0$, where $M_0$ is the rest mass of the star. Three branches of solutions are shown -- the trivial branch of solutions and two branches of scalarized solutions. Both branches with nonzero scalar field are energetically more favorable over the general relativistic solutions.}
	\label{fig:M_M0}
\end{figure}

Let us discuss now which one of the branches of solutions would realize in practice. For this purpose we plot the quantity $1-M_0/M$, that is related to the binding energy of the star, as a function of $M_0$ in Fig. \ref{fig:M_M0}, where $M_0$ is the rest mass of the star. For better visibility, only the trivial branch of solutions with $\varphi=0$ and two nontrivial branches characterized by scalar field which has no zeros for different values of $\lambda$ and $\beta$ are shown. The results for the other nontrivial branches where $\varphi(r)$ has no zeros is qualitatively the same. As one can see there is a cusp in the general relativistic case that is exactly the turning point in the $M(\rho_0)$ diagram and signals a change of stability. For the branch with $\lambda=20$ and $\beta=200$ no cusp exists since this branch is terminated because of violation of condition \eqref{eq:SolutionConstrain} before the mass in the $M(\rho_0)$ diagram reaches maximum. The branch with $\lambda=20$ and $\beta=400$ on the other hand has a cusp since the mass can reach a maximum in this case. What is important is that both nontrivial branches have higher binding energy (in terms of absolute value) and therefore they will be the preferred solutions over the pure general relativistic case. This means that, similar to the spontaneous scalarization of neutron stars in scalar-tensor theories of gravity, the trivial solution in stable and would realize in practice for lower central energy densities. At the bifurcation point this solution looses stability and the scalarized solutions become energetically more favorable. Of course at higher densities other nontrivial branches can appear characterized by $\varphi(r)$ which has one or more zeros, but our studies show that these branches are in general very short and that is why they deviate marginally from the pure general relativistic solutions. More over, they have lower binding energy (in terms of absolute value) than the solutions without zeros of $\varphi(r)$  and it is supposed that they are unstable.   

\begin{figure}[htb]
	\includegraphics[width=0.495\textwidth]{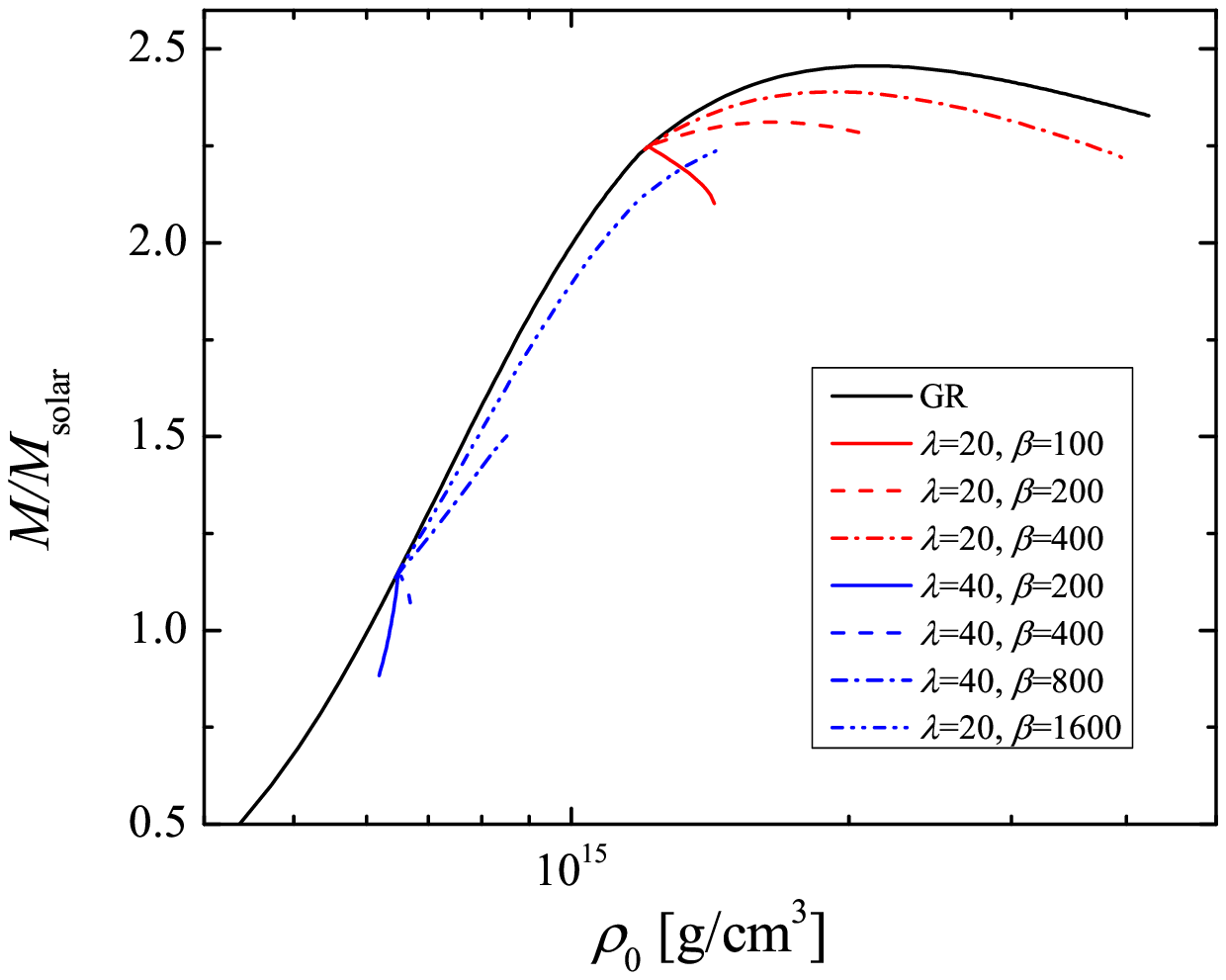}
	\includegraphics[width=0.495\textwidth]{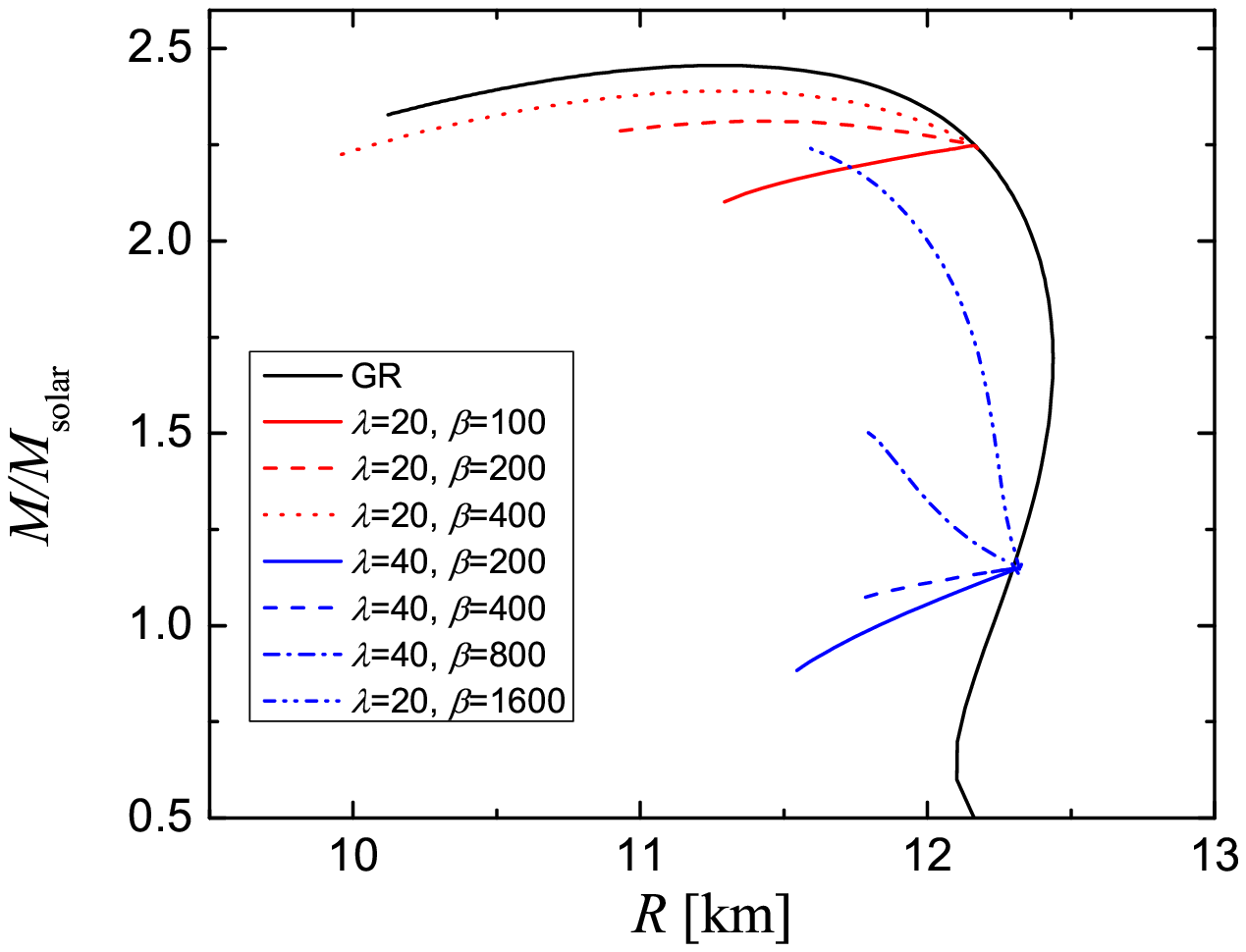}
	\caption{The mass as a function of the central energy density (left panel) and as a function of the radius (right panel) in the case of the coupling function \eqref{eq:coupling_function2}. The case of pure general relativity (trivial branch with $\varphi=0$) is shown as well as several scalarized branches with different values of $\lambda$ and $\beta$. Only solutions characterized by no zeros of the scalar field function $\varphi(r)$ are shown. }
	\label{fig:M_R_eps2}
\end{figure}

In order to have completeness of our results we explored the solutions with a little bit different coupling function, namely
\begin{equation}
f(\varphi)= - \frac{1}{2\beta} \left[\exp(-\beta\varphi^2)-1\right], \label{eq:coupling_function2}
\end{equation} 
i.e. we have just changed the sign of the original function, i.e. $\epsilon=1$. It is worth noting that  (\ref{eq:coupling_function2}) is just of the  type  coupling functions for which  the existence of scalarized black holes in ESTGB gravity was proven in our previous work \cite{Doneva_2017}.   
One of the main differences with the case discussed until now, i.e. with (\ref{eq:coupling_function}),  is that the parameter $\epsilon$ in equation \eqref{eq:Potential} changes sign. This leads to the fact that the negative minimum of the potential is much smaller, while the positive part is larger as discussed also in \cite{Silva_2017}. Since the negative part of the potential is controlled also by the coupling parameter $\lambda$, scalarized solutions develop in general for larger values of $\lambda$. For example, for the $\lambda=10$ case presented in the figures above, there is no scalarized solution when one uses the coupling function \eqref{eq:coupling_function2}. The mass as a function of the central energy density and the radius in this case is given in Fig. \ref{fig:M_R_eps2}. One can compare the two figures \ref{fig:M_R_eps} and \ref{fig:M_R_eps2} in the case of $\lambda=20$. If one uses the coupling function \eqref{eq:coupling_function} the bifurcatioin point happens at considerably smaller central energy densities. Another interesting difference is that even for very large $\beta$ where the solutions are close to the pure general relativistic case, the sequences terminate because of violation of condition \eqref{eq:SolutionConstrain} at much smaller central energy densities compared to the case with coupling function \eqref{eq:coupling_function}. Otherwise, the results in both cases are qualitative very similar.

\section{Conclusions}
In the current paper we have constructed realistic scalarized neutron star solutions in extended scalar-tensor-Gauss-Bonnet gravity where the source of the nontrivial scalar field is the curvature of the spacetime in contrast to the scalar-tensor theories where the source is the neutron star matter. The results show that for certain regions of the parameter space more than one neutron star solutions is present -- in addition to the trivial solution with zero scalar field (the analog to the pure general relativistic case) one or more scalarized solutions, characterized by the number of zeros of the scalar field, can exist. In our calculations we have used two different scalar field coupling functions  which permit such scalarization and both of them lead to qualitatively similar results.

In order to facilitate the search of the bifurcation points where new branches of solutions with nontrivial scalar field bifurcate from the trivial branch, we have examined the stability of the trivial solutions. For this purpose we have studied the perturbation of the pure general relativistic solution within the framework of the ESTBG gravity. It turns out that the trivial solution is stable up to some critical central energy density determined by the scalar field coupling parameter $\lambda$ and the coupling function. At the point of change of stability it is expected that new nontrivial branch of solutions bifurcate from the trivial one and we have explicitly constructed such neutron star solutions by solving the reduced field equations. 

The analysis of the field equations show that the neutron star solutions can not exist for arbitrary values of the parameters. Instead, one can derive a fourth order algebraic equation for one of the coefficients in the expansion of the metric close to the center of the star and solutions exit only when real roots of this equation exist. This leads to the fact that the branches of nontrivial solutions are terminated at some value of the central energy density before reaching a maximum. 

Our results show that the maximum mass for the sequences of neutron stars with nonzero $\varphi$ decreases with respect to pure general relativity. Therefore, in order to meet the requirement that the maximum mass of the neutron star sequences should be above two solar masses, one has to impose constraints on the parameters of the theory. Such a constraint would of course heavily depend on the particular choice of the equation of state and the scalar field coupling function.

At the end we have studied the binding energy of the obtained solutions. It turns out that the scalarized neutron stars characterized by no zeros of the scalar field have higher binding energy (in terms of absolute value) compared to the pure general relativistic case and that is why it is expected that they are stable and energetically more favorable. The solutions characterized by scalar field with one or more zeros have lower binding energy with respect to the solutions without zeros of $\varphi(r)$ and it is expected that they are unstable.

\section*{Acknowledgements}
DD would like to thank the European Social Fund, the Ministry of Science, Research and the Arts Baden-W\"urttemberg for the support. DD is indebted to the Baden-Württemberg Stiftung for the financial support of this research project by the Eliteprogramme for Postdocs. The support by the Sofia University Research Fund under Grant No 3258/2017  and COST Actions MP1304, CA15117, CA16104   is also gratefully acknowledged.

%%%%%%%%%%%%%%%%%%%%%%%%%%%%%%%%%%%%%%%%%%%%%%%%%%%%%%%%%%%%%%%%%%%%%%%%%%%%%%%


\begin{thebibliography}{99}

\bibitem{Berti_2015} E. Berti, E. Barausse, V. Cardoso, L. Gualtieri, P. Pani, U. Sperhake, L. C. Stein, N. Wex, K. Yagi, T. Baker, C. P. Burgess, F. S. Coelho, D. Doneva, A. De Felice, P. G. Ferreira, P. C. C. Freire, J. Healy, C. Herdeiro, M. Horbatsch, B. Kleihaus, A. Klein, K. Kokkotas, J. Kunz, P. Laguna, R. N. Lang, T. G. F. Li, T. Littenberg, A. Matas, S. Mirshekari, H. Okawa, E. Radu, R. O?Shaughnessy, B. S.
Sathyaprakash, C. Van Den Broeck, H. A. Winther, H. Witek, M. Emad Aghili, J. Alsing, B. Bolen, L. Bombelli, S. Caudill, L. Chen,
J. C. Degollado, R. Fujita, C. Gao, D. Gerosa, S. Kamali, H. O. Silva, J. G. Rosa, L. Sadeghian, M. Sampaio, H. Sotani, and M. Zilhao,
Classical and Quantum Gravity 32, 243001 (2015); arXiv:1501.07274 [gr-qc].

\bibitem{Pani_2011} P. Pani, E. Berti, V. Cardoso and J. Read, Phys. Rev. D84, 104035 (2011); [arXiv: 1109.0928].

\bibitem{Yunes_2011} N. Yunes and L. C. Stein, Phys. Rev. D 83, 104002 (2011); [arXiv: 1101.2921].

\bibitem{Pani_2011a} P. Pani, C. F. B. Macedo, L. C. B. Crispino and V. Cardoso, Phys. Rev. D 84, 087501 (2011);[arXiv: 1109.3996].

\bibitem{Pani_2011b} P. Pani, E. Berti, V. Cardoso and J. Read, Phys. Rev. D84, 104035 (2011), [arXiv: 1109.0928].

\bibitem{Slasedo_2016} J. L. Blzquez-Salcedo, L. M. Gonzlez-Romero, J. Kunz, S. Mojica and F. Navarro-Lerida, Phys.
Rev. D93 (2016) 024052, [arXiv: 1511.03960].

\bibitem{Kleinhaus_2014}  B. Kleihaus, J. Kunz and S. Mojica, Phys.Rev. D90 (2014) 061501, [arXiv: 1407.6884].

\bibitem{Kleinhaus_2016} B. Kleihaus, J. Kunz, S. Mojica and M. Zagermann, Phys. Rev. D93 (2016) 064077, [arXiv:
1601.05583].


\bibitem{Doneva_2017} D. D. Doneva and S. S. Yazadjiev, {\it New Gauss-Bonnet black holes with curvature induced scalarization in the extended scalar-tensor theories}, arXiv:1711.01187
[gr-qc].

\bibitem{Silva_2017} H. O. Silva, J. Sakstein, L. Gualtieri, T. P. Sotiriou
and E. Berti, {\it Spontaneous scalarization of black holes and compact stars from a Gauss-Bonnet coupling}, arXiv:1711.02080 [gr-qc].

\bibitem{Antoiou_2017a} G. Antoniou, A. Bakopoulos, P. Kanti, {\it Black-Hole Solutions with Scalar Hair in Einstein-Scalar-Gauss-Bonnet Theories}, 	arXiv:1711.07431 [hep-th].

\bibitem{Antoiou_2017b} G. Antoniou, A. Bakopoulos, P. Kanti, {\it Evasion of No-Hair Theorems in Gauss-Bonnet Theories},
arXiv:1711.03390 [hep-th]


\bibitem{Damour_1993} T. Damour and G. Esposito-Farese, Phys. Rev. Lett. 70, 2220 (1993).

\bibitem{Muther1987} H. Muther, M. Prakash, and T. L. Ainsworth, Phys. Lett. B 199, 469 (1987).

\bibitem{Read2009} J. Read, B. Lackey, B. Owen, and J. Friedman, Phys. Rev. D 79, 124032 (2009).

\bibitem{LIGO_NSMerger} B. P. Abbott et al., Phys. Rev. Lett. 119, 161101.
\end{thebibliography}
\end{document}